\documentstyle[psfig]{l-aa}
\begin{document}
\thesaurus{section(08.16.7; 13.25.5)}
\title{ Fine phase resolved spectroscopy of the X-ray emission of the 
Crab pulsar (PSR B0531+21) observed with BeppoSAX}
\author{E. Massaro\inst{1}, G. Cusumano\inst{2}, M. Litterio\inst{1}, 
T. Mineo\inst{2}
}
\institute{Istituto Astronomico, Universit\'a di Roma "La Sapienza",
Unit\`a GIFCO--CNR,via G.M. Lancisi 29, I-00161 Roma, ITALY \and
Istituto di Fisica Cosmica ed Applicazioni all'Informatica, CNR, via U. La 
Malfa 153, I-90146 Palermo, ITALY 
}
\offprints{E. Massaro}
\date{Received ; Accepted }
\maketitle
\markboth{E. Massaro et al.: Phase resolved spectroscopy of the X-ray emission 
of Crab pulsar }
{E. Massaro et al.: Phase resolved spectroscopy of the X-ray emission  
of Crab pulsar }

\begin{abstract}

The Crab Nebula and the Pulsar have been pointed several times by the 
BeppoSAX satellite.  
We analysed all these observations, summed together, to study the main spectral
properties with a good phase resolution. 
A new accurate estimate of the hydrogen column
density in the Crab direction  $N_{\rm H}$=3.23$\times$10$^{21}$
cm$^{-2}$ is given as derived from the analysis
of the off-pulse 
emission at low energies. 
We studied the changes of the spectral index with the pulse phase 
and showed that in the interpeak region it is systematically harder than in the 
main peaks of about 0.4.
We observed also a significant
spectral steepening of the pulsed emission and showed that this distribution 
can be represented by a unique law with a linear variable log slope and
apply it to the spectrum of the first peak.
A two component model 
is proposed to explain the phase evolution of the spectrum. One  component 
has the same profile observed in the optical while the other 
presents a maximum at the phase 0.4 with essentially the
 spectrum  
of the interpeak region. 
Two possible interpretations for the origin of the latter component are discussed.

\end{abstract}

\section{Introduction}

It is well known since early X-ray observations that the spectral 
energy distribution of the Crab pulsar (PSR B0531+21) is changing along 
with its double peak pulse profile. 
The intensity ratio of the second to the first peak increases from the 
soft X-rays up to the MeV region where it reaches the highest value 
(see the data archive of Massaro, Feroci \& Matt 1997). 
The origin of this effect is not understood up to now.
Pulse shapes have been computed for different accelerator sites: polar 
caps (Sturner \& Dermer 1994, Daugherty \& Harding 1996, Miyazaki \& Takahara 
1997), outer gaps (Chiang \& Romani 1994, Romani \& Yadigaroglu 1995) and 
closed magnetosphere (Eastlund, Miller \& Michel 1997). 
These theoretical 
profiles, in some cases symmetric, do not match the energy dependence 
of the peak intensity ratio and cannot explain the spectral changes across 
the pulse profile.

A first phase resolved X-ray spectroscopy of the Crab pulsar was performed 
by Pravdo \& Serlemitsos (1981) who used the data of one day long observation 
of OSO 8. 
More recent data, with a much finer phase resolution, were presented by
Pravdo, Angelini \& Harding (1997, hereafter PAH) who analysed the
RossiXTE data between 5 and 250 keV.

The Italian-Dutch satellite BeppoSAX observed the Crab Nebula and Pulsar 
in 1996 during the Science Verification Phase (Mineo et al. 1997). 
From that epoch to the end of 1998 this source was 
pointed several times because it is used for a calibration check of the 
Narrow Field Instruments (hereafter NFI; for a detailed description of 
the scientific payload of BeppoSAX see Boella et al. 1997). 
All these observations, when summed together, provide a high statistics 
data set which can be used for a very accurate phase resolved spectroscopy 
of the pulsed emission over an energy range wider than two orders of 
magnitude, from about 0.1 up to 300 keV. 
In this paper we present the results of an accurate spectral analysis 
that shows in detail the relevant hardening of the spectral slope in the 
interpeak region 
with respect to the two main peaks. 
Our analysis is substantially more complete than PAH because of
the wider energy band covered by BeppoSAX and the improved statistics
due to a longer total exposure.
We discuss, in particular, the spectral shape at energies lower than
4 keV and give a new accurate estimate of the absorbing column density in
the Crab direction. Furthermore, we show that a single power law does 
not give an acceptable spectral fit over the entire range and 
that a better description can be achived using a countinously 
steepening spectral energy distribution.
Finally,  we also discuss that the change with phase of the spectral slope 
can be explained by the superposition of 
two emission components, likely originated in different regions of 
the magnetosphere.

%
%  figure 1:  one column, 7 cm high 
%

\begin{figure}
\centerline{
\hbox{
\psfig{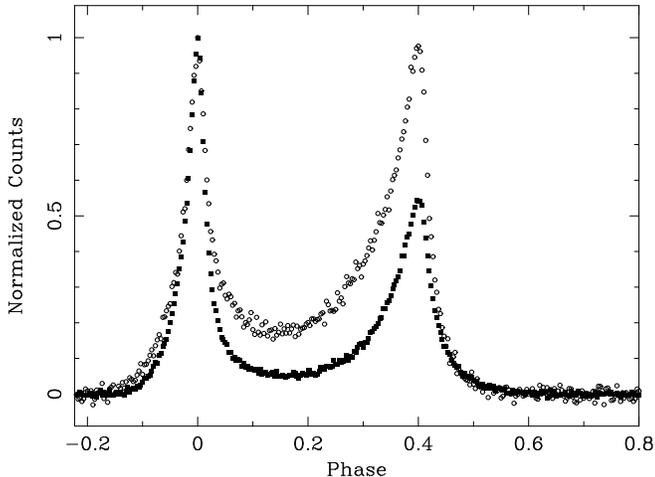}
}}
\caption{Two summed X-ray pulse profiles of Crab  in 300 phase channels 
in the energy ranges 1.5--5 keV  (filled squares) 
and 100--300 keV (open circles). 
Both profiles are normalized to unity at
the maximum of the first peak, after the subtraction of the off-pulse 
(0.6--0.83) constant level. 
Note the excess in the leading edge of first peak at higher energies.
}
\label{fig1}
\end{figure}

\section{ Observations and Data Reduction}

After the first observation of the Crab (1996 August - September),
BeppoSAX pointed this source several times and collected a large 
amount of good quality data. The epochs of the 
pointings and the net exposure times for each NFI are given in Table 1. 
We recall, however, that after May 1997 the MECS operated with reduced 
sensitivity because of the failure of one of the three detector units. 
With respect to Mineo et al. (1997) the data set has increased by factors 
ranging from 2.1 for the HPGSPC to about 4.9 for the LECS. 
Such great improvement of the statistics allows now to perform a fine 
resolved spectral analysis even in the 
interpeak region,
where the pulsed signal is low, particularly at energies below a few keV.

\begin{table}
\caption{ The BeppoSAX NFIs pointing epochs and the net exposure times of
 the Crab Pulsar.}
\begin{flushleft}
\begin{tabular}{lrrrr}
\hline
\multicolumn{1}{c}{Observation } & \multicolumn{4}{c}{Exposure Times (s)} \\
\multicolumn{1}{c}{Date} & \multicolumn{1}{c}{LECS} & \multicolumn{1}{c}{MECS} &
 \multicolumn{1}{c}{HPGSPC}  & 
 \multicolumn{1}{c}{PDS} \\
% &keV  0.1--4 & 1.6--10   & 10--34   & 15--300\\
\hline
31~Aug~1996     & --      &  --     &  28\,384 & 27\,749 \\
~6~Sep~1996     & 5\,733  & 33\,482 & --       &  --   \\
30~Sep~1996     & --      &  7\,874 &  --      & 6\,305  \\
11~Apr~1997     & 1\,257  & 18\,551 &  9\,344  &  9\,294 \\
~8~Oct~1997$^a$ & 12\,096 & 30\,769 &  9\,158  & 18\,622 \\
~6~Apr~1998$^a$ & 8\,730  & 28\,694 &  12\,732 & 13\,105 \\
13~Oct~1998     & --      &  --     &   --     & 14\,916 \\
%~9~Mar~1999    & --      &  --     &   --     & 10\,797  \\
\hline
%Total exposure      & 27\,816 & 119\,370 & 59\,618 & 100\,788  \\
Total exposure      & 27\,816 & 119\,370 & 59\,618 & 89\,991  \\
\hline 
\\
\end{tabular}

$^a$ MECS operated with two detectors after May 1997.
\end{flushleft}
\end{table}

For the imaging instruments we selected all the events within circular
regions centered at the source position and having radii of 4' 
(MECS) and 8' (LECS). This choice corresponds 
to use a percentage of about 90 \% of the total source signal in 
both instruments, but it allows to 
apply the best tested spectral response matrices. 

Phase histograms of the Crab pulsar were evaluated for each NFI and
each pointing using the period folding technique. 
The UTC arrival times of all selected events were converted to the
Solar System Barycentre with the DE200 ephemeris.
The values of $P$ and $\dot P$, for each observation epoch were derived
from the Jodrell Bank Crab Pulsar Monthly Ephemeris 
(http://www.jb.man.ac.uk/). 
We constructed a large set of 300 bin phase histograms for every 
energy channel of each NFI.
The zero phase was fixed at the centre of the first peak, evaluated by 
means of gaussian fits. 
All the histograms for the same energy channel of each NFI were 
then added together.
Before this operation we verified that all the profiles of the various
observation epochs had fully compatible shapes and therefore similar  
folding accuracy.
Two examples of summed pulse profiles for the MECS (1.5--5 keV) and PDS 
(100--300 keV), after the subtraction of a constant off--pulse signal,
taken in the phase interval 0.6--0.83, and normalized to unity at the 
maximum of the first peak, are shown in Fig. 1. We stress, in particular, 
the improved S/N ratio of the data at energies greater than 100 keV, in 
comparison with those of PAH.

For the phase resolved spectral analysis we subtracted the mean off-pulse level
from the content of  each phase bin. Spectra were accumulated in phase 
intervals having a minimum width of 0.00667 (= 1/150).
Whenever the count number is not
high, in the interpeak region and in some energy ranges, 
wider phase intervals were taken in order to reduce 
the statistical uncertainty in  estimating  the spectral parameters.
Furthermore, energy channels were rebinned up to have
a minimum content of 20 counts.

%
%  figure 2:  one column, 7 cm high 
%

\begin{figure}
\centerline{
\hbox{
\psfig{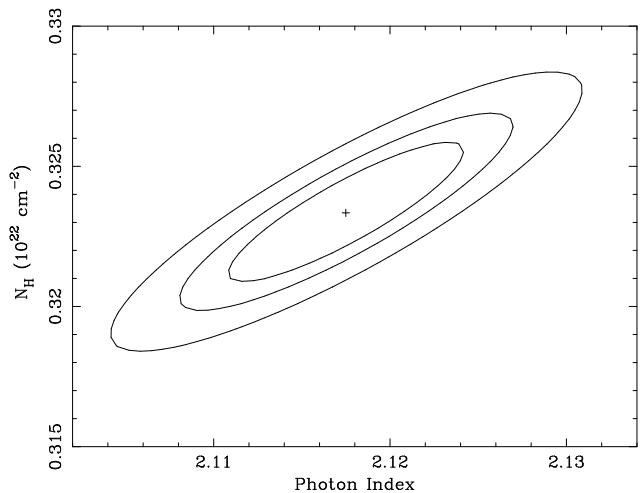}
}}
\caption{Contour plot for the column density vs the power law spectral index
derived from the LECS off-pulse spectrum  
(0.1 - 4 keV). The three contours correspond to 68, 90 and 99 \%
confidence levels; the cross is relative to the best fit values of 
the two parameters.
}
\label{fig2}
\end{figure}

The response matrices of all the NFIs used in our analysis are those 
relative to the November 1998 release.
They are based on ground
calibrations and MonteCarlo simulations and have been tested in flight with several
well known X-ray sources (among which Cassiopeia A, 3C\,273, NGC\,4151,
Vela X--1, Centaurus X--3, and the Crab Nebula itself). 
None of the NFI response matrices were  adjusted
in order to fit the Crab Nebula spectrum with a pre-defined power law, 
but this source was used as the other calibration sources to
verify the goodness of the on-ground calibrations.
(we thank F. Frontera, A. Parmar and A. Santangelo for information on this 
subject).

In the following we will shortly refer P1 and P2 for the first and second 
peak, respectively, and Ip for the bridge region between them. We stress 
that a standard definition for their phase widths is not existing, and we
will conventionally adopt (-0.05,0.05) for P1 and (0.30,0.44) for P2. In any
case, other phase intervals, when considered in our analysis, are always
indicated in the text.

\section{ X-ray Spectral Properties }

\subsection{ The off-pulse spectrum and the N$_H$ estimate }

An important quantity necessary for the spectral analysis at energies
lower than a few keV is the intervening column density $N_{\rm H}$. 
Although the Crab is one of the best studied X-ray sources this quantity
is not well known. Several significantly different estimates based on
various methods, ranging from 1.6 $\times$10$^{21}$ cm$^{-2}$ (Clark 1965) 
to 3.4 $\times$10$^{21}$ cm$^{-2}$ (Fritz et al. 1976), are reported in 
the literature.
Recent estimates of the interstellar reddening agree for a mean $E(B-V)$ 
equal to 0.53 $\pm$ 0.04 (Blair et al. 1992, Gull et al. 1998). Assuming
$R=A_V/E(B-V)$=3.09 (Rieke \& Lebofsky 1985) and the $N_{\rm H}$--to--$A_V$
conversion by  Predehl \& Schmitt (1995), we obtain the value of 
2.9$\times$10$^{21}$ cm$^{-2}$. If the conversion factor estimated
by Gorenstein (1975), also reported by Zombeck (1990),
is used, the column density increases up to 3.4 $\times$10$^{21}$ 
cm$^{-2}$ 

We computed a new estimate of this quantity from  BeppoSAX data. To 
this aim we analysed the spectrum of the off-pulse region assumed 
representative of the nebular emission, which is well described by 
a single power law in a wide energy interval, practically covering all
the entire BeppoSAX range, as it will be discussed in the following.
A fit with an absorbed power law of the LECS data 
for energies between 0.1 and 4 keV gives a spectral index of 2.117 $\pm$ 
0.004 and $N_{\rm H}$= (3.23 $\pm$ 0.02) $\times$10$^{21}$ cm$^{-2}$ with a 
reduced $\chi^2$=1.3 (364 d.o.f.). 
The contour plots for these two parameters are shown in Fig. 2; we can see
that column densities smaller than 3.18 $\times$10$^{21}$ cm$^{-2}$ are
evidently not compatible with a simple power law.  If
the value $N_{\rm H}$=2.9 $\times$10$^{21}$ cm$^{-2}$ is used the spectral 
index lowers to 2.04, but the reduced $\chi^2$ reaches the largely 
unacceptable value of 2.5. 
Furthermore, the best fit spectral index of the MECS data between 4 and 7 keV is 
2.117 $\pm$ 0.003 independently of the adopted $N_{\rm H}$,
coincident with that given above. 

A value of $N_{\rm H}$ equal to 3.23$\times$10$^{21}$ cm$^{-2}$ corresponds to 
a conversion factor with $E(B-V)$ of 6.09$\times$10$^{21}$ cm$^{-2}$ 
mag$^{-1}$, about 20\% greater than that of Predehl \& Schmitt (1995), 
but closer to the one by Gorenstein (1975). 

Finally we stress that $N_{\rm H}$ values of 3.0 - 3.2 $\times$10$^{21}$ 
cm$^{-2}$ were also found in the analysis of some ASCA observations of the
Crab (Fukazawa, Ishida \& Ebisawa 1997).
We are, therefore, confident that our result is likely the best available 
estimate of the actual column density and used it in  the 
spectral analysis.

\subsection{ Phase Evolution of the Spectral Indices } 

Phase dependent photon indices for single power laws were computed 
from the data of each instrument covering different energy ranges: LECS 
(0.1-4.0 keV), MECS (1.6-10 keV), HPGSPC (10-34 keV) and PDS (15-300 keV). 
The results are shown in the four panels of Fig. 3.
The reduced $\chi^2$ values are generally acceptable: they span the range 
between 0.8 and 1.25
with only four values over 127 fits greater than 1.25.

We computed also the photon indices for two separate energy ranges of PDS,
16-80 keV and 80-300 keV, and the results are shown in Fig. 4. 
The same phase dependence is clearly apparent in each plot: the
feature with the softest spectrum is P1, while the middle of the Ip 
is the hardest; the spectral index difference $\Delta \alpha$ is of
the order of 0.30 - 0.45. 
Spectral indices are clearly increasing with energy over all the phase interval.
In particular, that of P1 changes from 1.6 in the LECS range 
(Fig. 3a) to 2.3 at higher energies (Fig. 4b). The latest value is 
in good agreement 
with some other recent results above 100 keV like those of FIGARO II 
(Massaro et al. 1998) and CGRO (Ulmer et al. 1995).
Note also that in  Fig. 3c,d the central bins of P1
have a softer spectrum than the wings.

%
%  figure 3:  one column, 7 cm high 
%

\begin{figure}
\centerline{
\hbox{
\psfig{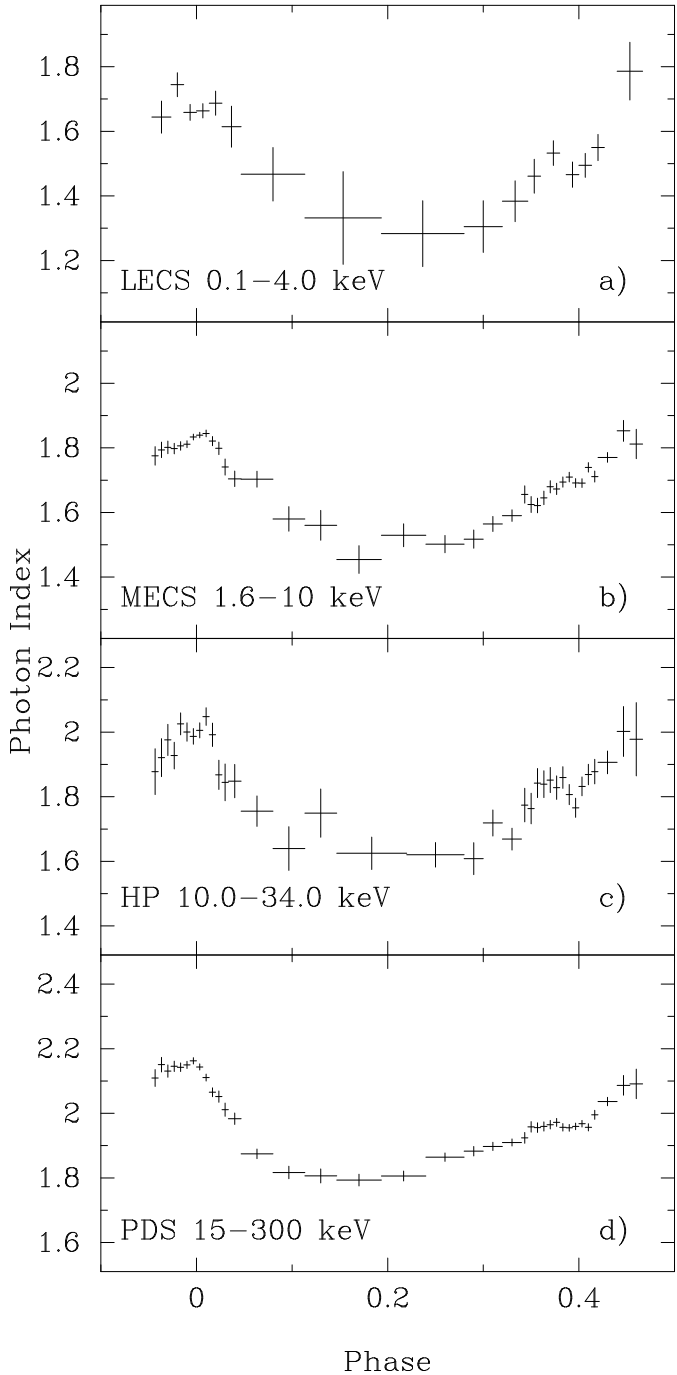}
}}
\caption{The phase resolved spectral index in the energy ranges of the four NFIs:
LECS 0.1--4.0 keV panel a), MECS 2--10 keV panel b), HPGSPC 10.0 --34.0 keV
panel c), PDS 15--300 keV panel d).
}
\label{fig3}
\end{figure}

PAH reported similar results obtained with the PCA (5~-~60 keV) and 
HEXTE (16~-~250 keV) instruments on board RossiXTE. 
The phase evolution of the spectral index is nearly coincident with that 
found by us, but the spectrum steepening is not so clear as in our results.
In particular, the P1 indices in the PCA and HEXTE data are 1.95 and 2.05.
The latter value agrees with ours in the HPGSPC range (Fig. 3c) but it is
smaller than that of the PDS (Fig 3d, Fig. 4); the former is
intermediate between those of MECS and HPGSPC, as expected by the different
energy ranges and by  the PCA lower threshold of 5 keV.
The spectral indices of central Ip region are also in agreement with 
ours: the PCA value (1.62) coincides with to that of the HPGSPC, 
whereas the HEXTE value (1.75) is a bit smaller than ours, but
the difference is not significant being within the HEXTE statistical 
uncertainty.

\subsection{ The spectrum of the first peak }

It is already known that the spectral energy distribution of the total
pulsed emission from Crab is continuously steepening from the optical
frequencies to $\gamma$-rays with a maximum in the $\nu F_{\nu}$ plot
in the MeV range (see, for instance, the review by Thompson et al. 1997). 
This distribution, however, is affected by the strong increase of P2 intensity
with respect to P1. 

%
%  figure 4:  one column, 7 cm high 
%

\begin{figure}
\centerline{
\hbox{
\psfig{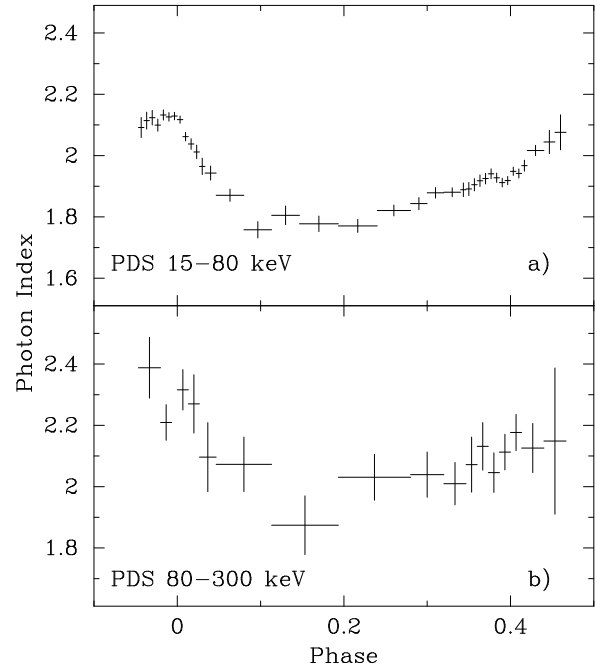}
}}
\caption{The phase resolved spectral indices in two PDS energy ranges:
16-80 keV (panel a) and 80-300 keV (panel b). 
}
\label{fig4}
\end{figure}

As stated above,
from the plots of Figs. 3 and 4 we see that the spectral indices 
in the various NFIs' energy ranges  are significantly different, and
therefore we expect that a single power law does not give a satisfactory 
representation of the spectral distribution over the entire 
BeppoSAX (0.1 - 300 keV)  range. 
We studied, in particular, the spectrum of P1 which is the feature with the
highest S/N ratio.
We selected the events in a small phase interval 0.02667 wide, 
corresponding to eight bins over 300, centred at the maximum (zero phase).
The reason for this choice, instead of the entire phase range of P1,
is explained by the model described in Sect. 4. In this way we take only
the bins where the P1 signal is stronger and minimize the possible
contribution of a much harder component.

The best fit of a single power law obtained by taking the data of all 
the four instruments at the same time gives the not acceptable reduced 
$\chi^2$ value of 3.5 (748 d.o.f.). 
Better fits are obtained when the NFIs are considered
separately (Fig. 5).
 This great variation is not apparent for the 
corresponding indices of the off-pulse (or nebular) spectrum, whose
values lie all in the narrow interval 2.12 - 2.16, with a typical 
statistical uncertainties of about 0.01 or less (Cusumano et al. 1998).
The small but significant differences found among the fits of the various 
NFIs are likely due to systematic effect.

%
%  figure 5:  one column, 7 cm high 
%

\begin{figure}
\centerline{
\hbox{
\psfig{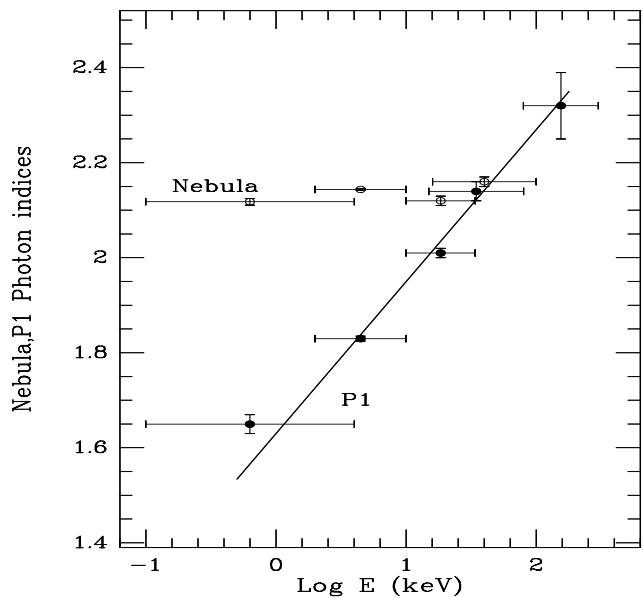}
}}
\caption{Comparison of the photon indices of P1 (filled circles) 
with that of the nebula (open circles).
The straight line corresponds to Eq. (2) when the best fit values of the
parameters $a$ and $b$ are used.
}
\label{fig5}
\end{figure}

%
%  figure 6:  one column, 7 cm high 
%

\begin{figure}
\centerline{
\hbox{
\psfig{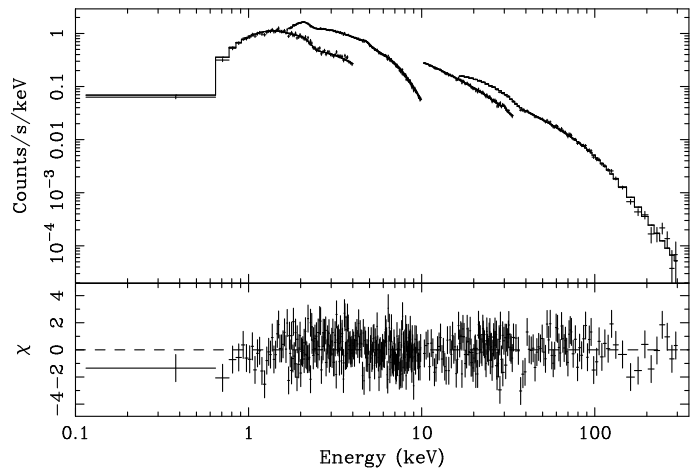}
}}
\caption{The spectral fit of P1 using the law given in Eq. (1) over the
wide energy range 0.1~-~300 keV covered by the four NFIs. 
}
\label{fig6}
\end{figure}

We tried then to represent the actual X-ray spectral distribution of P1
central bins with a continuously steepening law.
The simplest model is likely the one having a linear dependence of the 
spectral slope upon the logarithm of the photon energy $E$:

$$
F(E) = K (E/E_0)^{-(a + b~Log~(E/E_0))}\,\,,
\eqno(1)
$$

\noindent
with only three free parameters because we fixed $E_0$= 1 keV.
We stress that the spectral indices shown in Fig.3 indicate that
a similar law should hold also at other phases; a more general model
should so include the phase dependence of $K, a, b$.
When Eq. (1) is fitted to the data of all NFIs selected as above in 
the phase interval (-0.0133,~0.0133), with the $N_{\rm H}$
value given in Sect. 3.1, a much better agreement is achieved. 
The reduced $\chi^2$ lowers to 1.119 (747 d.o.f.) and the
best fit values of the three parameters are: $a$~=~1.63$\pm$0.08, 
$b$~=~0.16$\pm$0.04 and $K$~=~(8.87$\pm$0.07)$\times$10$^{-2}$ ph 
cm$^{-2}$ s$^{-1}$. 
The fitted spectrum and the residuals are shown in Fig. 6.
The $\chi^2$ value, although acceptable, is two standard deviations
greater than the expected one; this could be due to the occurrence
of very small residual systematic effects that, with such a bright 
source can give a not negligible contribution to the $\chi^2$.

The energy dependent photon spectral index $\alpha(E)$ can be evaluated
by the log derivative of Eq.(1): 

$$
\alpha(E) = a + 2~b~Log~(E/E_0)\,\,,
\eqno(2)
$$

\noindent
that corresponds to the straight line plotted in Fig. 5. 

We stress that the spectral distribution given by Eq.(1) with the best 
fit values of the parameters, is in principle valid only in the BeppoSAX
range, where they were evaluated.
Nevertheless, it reproduces the observed spectral shape when
extrapolated to the UV-optical-IR range: the resulting energy spectral 
distribution is nearly flat at the photon energy of about 3 eV, corresponding
to the photometric B band, and steepens toward the UV, in a good agreement 
with the results by Percival et al. (1993) and Gull et al. (1998).
The extrapolated flux, however, is smaller than the observed one.
The extrapolation of Eq.(2) at $\gamma$-ray energies ($>$ 30 MeV) 
gives a photon index greater than 2.8, higher than the observed one 
equal to 2.05$\pm$0.03 (Fierro 1995). 
This value is also smaller than that found by us for the PDS data and
implies that the P1 spectrum flattens in the MeV region.

\section{ A two component model }

In order to explain the behaviour of the spectral index with the phase, 
PAH proposed a scenario where the X-ray emission is originated by particles
accelerated in the polar cap region just above the neutron star surface.
The geometry of the emission beam is a hollow cone with the axis coaligned
to the magnetic one and the two main peaks correspond to the crossing
of the cone annulus by the line of sight. According to these authors the 
resulting spectrum is harder in the Ip where the emission is pure
curvature radiation not softened by the contribution from the $e^+~e^-$
cascade, more relevant close to the outer rim.
This interpretation, however, faces with some difficulties: the 
similar spectral behaviour of P2 and Ip 
and the spectral hardening of the P1 leading edge 
with respect to the central bins.

We followed a different approach and considered that the observed 
pulsed emission is due to the superposition of two components having 
different phase distributions and energy spectra.
The first component is assumed to have the same pulsed profile observed at 
optical frequencies (Smith et al. 1988, Percival et al. 1993), with P1 
much more prominent than P2 and a very low intensity in the Ip region. 
A similar shape is also observed at energies greater than 30 MeV
even if the profile of P2 is slightly different (Fierro et al. 1998)
suggesting that the electron populations emitting in both energy
ranges. We will refer to this component in the 
following as the optical one (shortly $C_{O}$). 
The second component (hereafter $C_{X}$) is assumed to have the greatest
relative intensity in the hard X~-~low energy $\gamma$ rays and it is,
therefore, responsible of the P2/P1 change. We derived its phase profile 
in order to reproduce the observed one when summed to $C_{O}$: we found that
$C_X$ increases monotonically up to the phase 0.4 and then has a sharp cut-off.
The main properties of the latter component were then estimated from the 
BeppoSAX data, essentially by means of a fitting procedure of several 
pulse profiles at different energies, as described in detail in the
following subsections.

%
% figure 7:  one column, 10 cm high 
%

\begin{figure}
\centerline{
\hbox{
\psfig{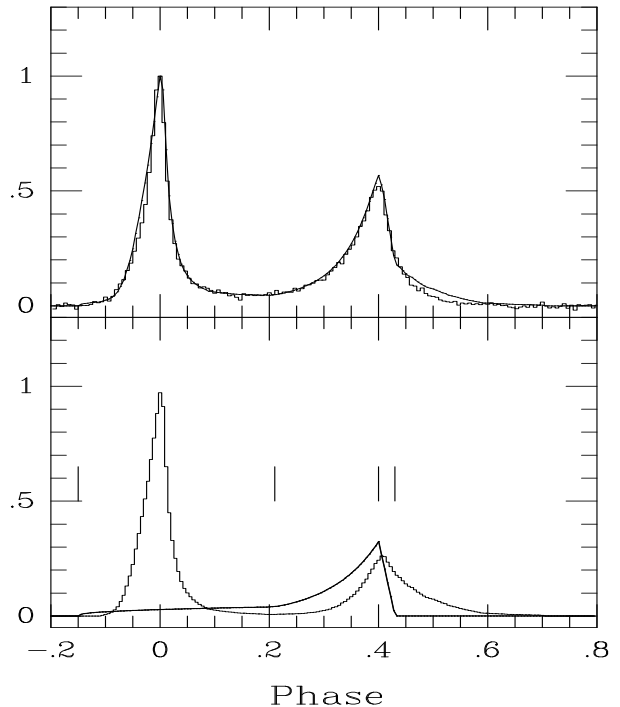}
}}
\caption{
Comparison between the observed pulse profile in the energy 
interval (1.0--1.95) keV and the two component model profile, computed
for the mean energy of 1.6 keV (upper panel). 
The phase 
distributions of $C_{O}$ and $C_X$ with the proper normalisation
are shown in the lower panel; the four vertical bars indicate the phases
from $f_1$ to $f_4$.
}
\label{fig7}
\end{figure}

\subsection{ The components' shapes}

The study of $C_X$ is easier if an analytical representation for it is used.
At the beginning we assumed that its phase distribution is substantially
independent of energy and therefore it is possible to write the $C_X$ 
intensity as the product of two functions one dependent only on
the photon energy $E$ 
and the other on the phase $f$:

$$
I_X(E,f)=Y(E)~g(f) \,\,,
\eqno(3)
$$

\noindent
The whole phase interval where $C_X$ is not zero is denoted by ($f_1$,$f_4$) 
and is divided by two inner points at the phases $f_2$ and $f_3$ into 
three segments: in ($f_1,f_2$) the $C_X$ shape is given by
a power law, in ($f_2,f_3$) by an exponential function, joining the 
first one at $f_2$ and finally a linear descending branch from $f_3$ 
to $f_4$ connects the maximum to the zero level of the off pulse. We have

$$
g(f)= \exp\{p (f_2-f_3)\}\left(\frac{f-f_1}{f_2-f_1}\right)^{s}\,\,,
\eqno(4a)
$$

\noindent
for $f_1<f<f_2$, and

$$
g(f)=\exp\{p(f-f_3) \}\,\,,
\eqno(4b)
$$

\noindent
for $f_2<f<f_3$.

We verified also if all the parameters of $g(f)$ are really independent 
of the photon energy. This assumption resulted quite correct for the 
phase interval boundaries, taken equal to $f_1=-0.15$, $f_2=0.21$, $f_3=0.40$, 
$f_4=0.43$ and the power law exponent $s=0.4$.
Note that with this choice $C_X$ is non zero also in the phase interval
of P1: it is necessary to explain the broadening of this peak with 
increasing energy and the softer spectrum of the peak centre with respect
to both wings.
As it will be described in sect. 4.2, we found that a better agreement 
with the data is obtained if also $p$ is allowed to vary: it changes
from about 11. in the keV range to 8. above 150 keV. The factorisation 
of Eq. (3), even if largely acceptable, does not give a very precise 
description of the pulse profiles, suggesting that the shape of $C_X$ 
cannot be considered completely independent of the energy.

%
% figure 8:  one column, 10 cm high 
%

\begin{figure}
\centerline{
\hbox{
\psfig{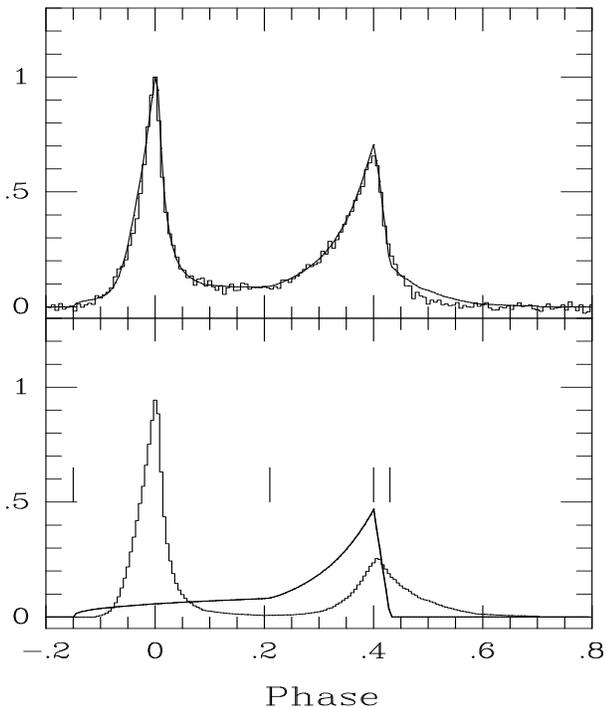}
}}
\caption{ The same as Fig. 7 but for the energy interval (8.0--10.0) keV
and a model pulse profile computed for the mean energy of 8.85 keV.}
\label{fig8}
\end{figure}

%
% figure 9:  one column, 10 cm high 
%

\begin{figure}
\centerline{
\hbox{
\psfig{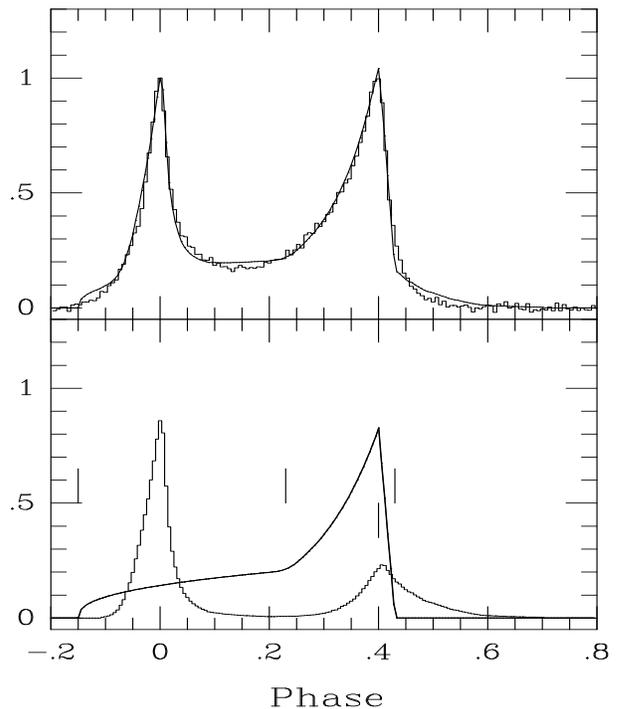}
}}
\caption{ The same as Fig. 7 but for the energy interval (57.--102.) keV
and a model pulse profile computed for the mean energy of 75.2 keV.}
\label{fig9}
\end{figure}

The estimates of $Y$ and $p$ were obtained by means of quadratic
minimization of the differences with the measured profiles in the 
($f_2$,$f_3$) interval only. The algorithm was slightly more complicated than 
usual because we worked
with profiles normalized to the P1 maximum and then $C_X$ contributes to
the normalization itself. For a given values of $E$ and $p$, $Y$ was 
estimated by the formula

$${
Y=\frac{\sum_i[(I_O(f_i)-I(f_i) (A I_O(f_i)-e^{\{p (f_i-f_3)\}})]}
{\sum_i[ (A I(f_i) - e^{\{p (f_i-f_3)\}})(A I_O(f_i)-e^{\{p (f_i-f_3)\}})]}}
\eqno(5)
$$

\noindent
where the sum is over all the considered phase bins, $I$ and 
$I_O$ indicate the observed X-ray and $C_{O}$ profiles, respectively, and

$$
A=g(0)=0.705~\exp\{-0.19~p \}\,\,.
\eqno(6)
$$

We also tried a model with a power law instead of the exponential 
function of Eq.(4b) but, because no improvement was found with
respect to the above formula, it will not be considered in  further
analyses.

\subsection{The total profile }

Three examples of pulse profiles computed using the two component model are
shown in the upper panels of Figures 7,8 and 9 together with  150 bin 
histograms of the BeppoSAX data for the three energy intervals (1.0-1.95),
(8.0-10.0), (57.-102.) keV. The lower panel 
of each figure shows  $C_{O}$ and $C_X$ separately: their 
relative intensities have been scaled in order to obtain a total value 
at the zero phase equal to unity. 
The values of $p$ and $Y$ used in the computations are those corresponding
to the mean energies weighted with the detected counts.
The agreement with the data is satisfactory in the whole energy 
range covered by BeppoSAX even if some small discrepancies are apparent.
For example, the leading edge of P1 and the trailing edge of P2 do not 
match well the data. These discrepancies, likely due either to
the analytical model of $C_X$ or to the optical profile itself, are
practically unavoidable unless one adopts a very {\it ad hoc} model.

In the upper panel of Fig.10  we plotted Log $Y$ {\it vs} Log $E$:
a single power law gives a good representation of the trend and the 
resulting exponent is 0.28. 
In Fig.10 (lower panel) the values of the coefficient $p$ are also shown.
Its monotonically decrease indicates 
that the leading edge 
of P2 becomes shallower and shallower with increasing energy.

A more careful analysis showed also that the Ip intensity is even better 
represented if $f_2$ is slightly changed with energy from the above value: the range 
remains narrow, from 0.20 at the lowest energies to 0.23 at the highest one, 
without significant differences of the $Y$ and $p$ values. These improvements,
however, are quite small and therefore only the shape parameter $p$ has to be 
considered significantly dependent upon energy.

%
%  figure 10:  one column, 10 cm high 
%

\begin{figure}
\centerline{
\hbox{
\psfig{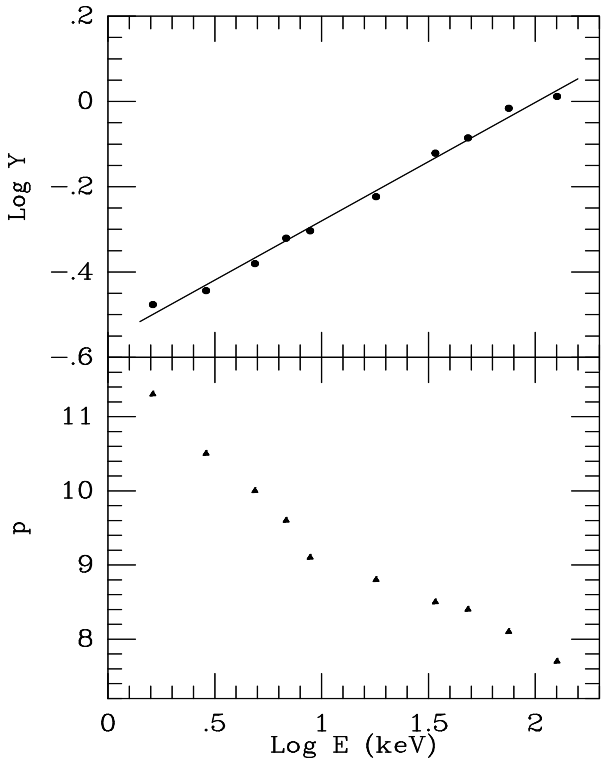}
}}
\caption{The energy dependence of the parameters $Y$ (upper panel) and 
$p$ (lower panel); the power law best fit of the former is also shown.
}
\label{fig10}
\end{figure}

\subsection{The spectral index}

The different spectral and phase distributions of the two components
would consistently reproduce the phase dependence of the spectral index 
described in Sect 3.2. This can be easily seen in Fig. 11 where our
data in the MECS band are compared with the model expectations.
The model photon index was simply computed from the ratio of the summed
intensities of $C_O$ and $C_X$ at the two extreme energy values:

$$
\alpha_{12}=\frac{Log(I(E_2)/I(E_1))}{Log(E_2/E_1)} + \alpha_O.
\eqno(7)
$$

Since we used normalised pulse profiles our slope is identically equal to 
zero at the centre of P1 and we added, therefore, the constant 
$\alpha_O$=1.83 to our results in order to match the measured values. 
The agreement between the observed points and the model is fully
satisfactory, in particular for the softness of the P1 core with respect 
to the wings.
The presence of a hard pedestal in the same phase interval of P1 is 
likely the simplest explanation of this asymmetry between the two peaks
that cannot simply explained by the hollow cone model of PAH.
Our results show that to reproduce the observations, the spectrum of 
this pedestal must be similar to that of Ip and P2 and therefore it 
is reasonable to consider them all  as belonging to the same component.

%
%  figure 11:  one column, 10 cm high 
%

\begin{figure}
\centerline{
\hbox{
\psfig{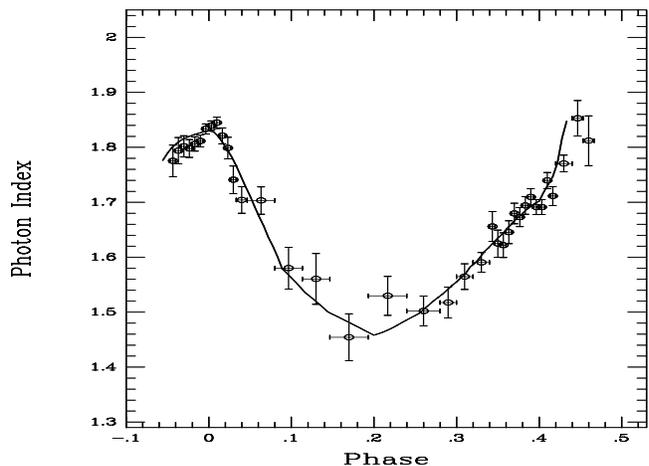}
}}
\caption{The photon index $vs$ phase in the MECS band expected from the 
two component model (solid line) compared with the results of Fig. 3b. 
}
\label{fig11}
\end{figure}

\section{ Discussion}

The BeppoSAX observations of the Crab pulsar have provided one of the 
best data set for a detailed study of the spectral and phase distributions 
over a wide X-ray energy interval. Using these data we derived the phase 
evolution of the spectral slope in several energy ranges. In particular,
we confirmed that the spectral index is not constant within the phase 
interval of P1 and that both the leading and trailing edges have harder 
spectra than the central bins.
Furthermore, we showed that the P1 spectral index changes from 1.65 to 2.3, 
for energies of about 0.5 keV to more than 200 keV. 
A law with a linearly varying spectral index gives a satisfactory 
representation of this behaviour.

The interpretation of all these findings in terms of a consistent physical
model is not simple. A first step in this direction is the understanding if
such complex behaviour is due to fact that we are observing
photons emitted in different regions of the magnetosphere and likely by
different emission mechanisms.
To this aim we showed that, at least as a first approximation, the shape of
X-ray pulse profiles of the Crab with the energy can be explained by 
the superposition of two components with different phase 
and energy distributions. 
The major success of this model is that it gives a very simple
explanation for the spectral index evolution with the phase. 
In particular, the soft core of P1 follows directly by the presence of 
 the harder $C_X$ pedestal.

>From the energy dependence of the relative component intensities we can
conclude that the X-ray spectrum of $C_X$ can be approximately described 
by a law similar to that of P1, but with a spectral index systematically
flatter by 0.28, so in the 2--10 keV range it would be equal to 1.55 and
 to about 2.0 around 200 keV. The $C_X$ spectrum should have 
a cut-off above this energy, likely above a few MeV, as indicated by COMPTEL 
data (Much et al. 1995). 
It is also possible that $C_X$ should be also detectable at lower energies,
down to optical frequencies, as suggested by the small residual flux in 
the Ip region. 

Open problems are the nature of the emission mechanisms and the locations 
in the magnetosphere for both these components. In the following we will
limit our discussion only to $C_X$, independently of the $C_O$ origin.
We propose two possible hypotheses; one based on
the spectral distribution and the other on the phase profile. A more
complete model of the Crab emission is beyond the aim of the present
paper.

Zhu and Ruderman (1997) have shown that a copious production 
$e^+ e^-$ pairs can occur in the closed magnetosphere not far
from the neutron star surface. Curvature photons, emitted by high 
energy particles leaving the acceleration site and moving toward the star 
along the field lines, make pairs where the transverse magnetic field 
becomes large enough to absorb them and perhaps colliding with X-ray 
photons from the star.
These pairs emit immediately synchrotron photons in the X-ray band and
the expected photon index is close to 1.5 if the emitting particles are 
only those of the first generation (Wang et al. 1998). 
Furthemore, these authors argue that the upper spectral cut-off should be 
in the MeV range. This spectral behaviour is quite similar to the one
we derived for $C_X$ and it is suggestive for such possible origin.
Wang et al. (1998) noticed also that hard X-ray components with similar 
photon indices are present in two other $\gamma$-ray pulsars (Geminga, 
PSR 1055-52) and associated it with this class of pulsars because
of the $e^+ e^-$ pairs production on closed field lines in the 
neighbouring of the neutron star. It is then natural that such a 
component could also be present in the Crab. More recently, Cheng \&
Zhang (1999) have shown that if more pair generation are considered the
photon index can change from 1.5 to 1.9. The computation of the phase
distribution, however, is a difficult task because the geometry of the
emission region and the radiation pattern can be quite complex.

%
%  figure 12:  one column, 10 cm high 
%

\begin{figure}
\centerline{
\hbox{
\psfig{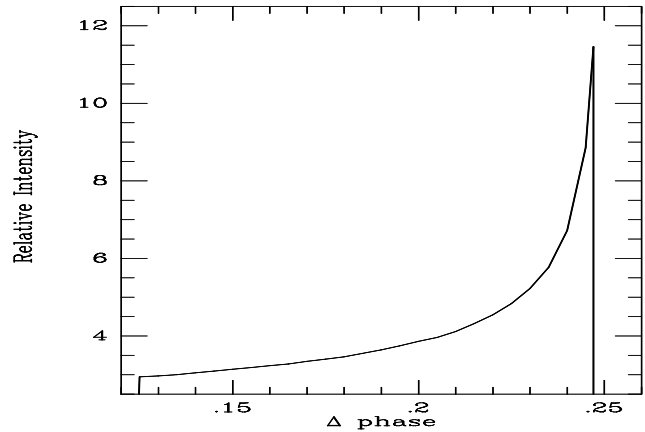}
}}
\caption{The expected phase distribution of photons due to the aberration 
shift and travel time effects. Photons are emitted radially in a plane 
perpendicular to the rotation axis from a region between 0.4 and 0.99 
of the light cylinder.
}
\label{fig11}
\end{figure}

Our other hypothesis on the $C_X$ origin is based upon its shape and
particularly the sharp cut-off beyond the phase 0.4. The observed phase
distribution is affected by the relativistic aberration which becomes
stronger and stronger approaching the light cylinder (Cheng, Ho \&
Ruderman 1986) and gives a phase shift of about 0.25. It is not difficult 
to show that the phase distribution of photons emitted radially in a 
region close to the light cylinder is similar to that of $C_X$. 
An example of this effect is plotted in Fig. 12, and the mathematical
justification is given in the Appendix.
Computations were carried out for a simple geometry approximating what was
expected from particles streaming along the outer gap field lines. 
We assumed that photons are emitted in a plane perpendicular to the 
rotation axis, in the radial direction with a $\delta$-like 
angular distribution in a region between 0.4 and 0.99 of the light cylinder. 
The resulting profile covers a phase interval  narrower than 
that we derived for $C_X$, about equal to 0.5.
This discrepancy can be solved by assuming an extended emitting region
and a not radial radiation pattern to achieve a combined phase width 
of about 0.3 - 0.4. 
This scenario is in conflict with the previous one because the emitting 
particles cannot be locally produced pairs this process being inhibited
by the relatively low magnetic field.

In conclusion we stress that the Crab pulsar would offer the interesting
possibility to observe at the same time radiation emitted by different
mechanisms in various regions of the magnetosphere. 
Fine resolved spectroscopy is likely the best way to study in  detail
these processes. It is important to extend it to higher energies, 
hopefully to the MeV range where the spectra of the components change, 
to achieve more information for determining the component parameters 
and understanding  the emission processes. The next scheduled INTEGRAL 
mission is one of the best opportunities to obtain very important 
 data in this range.

\begin{acknowledgements}
We are grateful to M. Ruderman for enlightening discussions on pulsar
models. This work has been partially supported by ASI - Agenzia Spaziale 
Italiana.
\end{acknowledgements}

\appendix{

\section{ Aberration effect in the phase distribution}

The observed phase of a photon is given by the sum of three terms
(Cheng, Ho \& Ruderman 1986): 
the emission phase $ \overline{\varphi }$, an aberration term $\varphi_a$
and a flight path difference term $\varphi_t$.
Suppose that at a point \( \overrightarrow{x}(\overline{r},
\overline{\theta },\overline{\varphi }) \) rotating with velocity
\( v=\beta c =\Omega \,\overline{r}\,\sin\overline{\theta } \) ($\Omega$ 
being the spinning velocity of the pulsar) a photon is 
emitted with 4-momentum \( k^{\mu }=\omega (1,\sin\theta \cos\varphi,
\sin\theta\sin\varphi,\cos\theta) \) in the corotating frame.
Then, assuming that photons are emitted along the radial direction 
($\theta=\overline{\theta}$, $\varphi=\overline{\varphi}$)
and neglecting any field bending, one gets the following simplified 
expressions:

\begin{equation}
\Delta \varphi_a=\arctan(\gamma \Omega \overline r /c)=\arctan (\beta\,\gamma 
\,\csc \theta )
\end{equation}

\begin{equation}
\Delta \varphi_t=(\Omega\overline r)/(\gamma c)={\frac{\beta\,\csc 
\theta }{\gamma }}
\end{equation}

\noindent
where $\gamma$ is the Lorentz factor of the local rotating frame.
Thus the total observed phase (in units of 2$\pi$ and taking $\varphi$=0) 
of the photon results to be:

\begin{equation}
\label{phi}
f = \frac{1}{2\pi} (\Delta \varphi_a+\Delta \varphi_t)
\end{equation}
%{}

The resulting range of $f$ is then from 0 to 0.25.
Suppose that photons are emitted in the interval 
$0<r_1\leq\overline r\leq r_2<c/\Omega\sin\overline{\theta}$
with some distribution function 
$dN(\overline r)=\psi(\overline r)d\overline r$. By using 
$ v=\Omega \,\overline{r}\,\sin\overline{\theta }$ and neglecting the 
$\overline{\theta}$ -- dependence of $r$ along the field -- line, 
this will become $dN(f)=n(f)df$, with:

\begin{equation}\label{nphi}
n(f)=\psi(\overline r)\frac{d\overline r}{df}\,\,.
\end{equation}
Thus, even if $\psi(\overline r)$ is constant (assumed unity) the 
distribution of the photons with the phase will be non--trivial.\par
After calculating the derivative of Eq.(\ref{phi}), the following expression 
for $n(\beta)=1/\partial_{\beta}f$ is found:

\begin{equation}\label{nbeta}
n(\overline r)=\frac{2\,\pi c}{\gamma \Omega sin\theta}\frac{1+
\beta^2\cot^2\theta}{1+(1-2\beta^2) (1+\beta^2\cot^2\theta)}\,\,.
\end{equation}

The final result for Eq.(\ref{nphi}) is obtained by combining (\ref{nbeta}) 
with (\ref{phi}) and can be computed numerically.

}


\begin{thebibliography}{}
\bibitem[1997]{BLL}  Boella G. Butler R.C., Perola G.C. et al. 1997, A\&AS, 
     122, 299 
\bibitem[1992]{BL}Blair W.P., Long K.S., Vancura O. et al. 1992, ApJ 399, 611
%\bibitem[1998]{CGZ}Cheng K.S., Gil J., Zhang L. 1998, ApJ 493, L35
\bibitem[1986]{CHR}Cheng K.S., Ho C., Ruderman M. 1986, ApJ 300, 500
\bibitem[1999]{CHZ}Cheng K.S., Zhang L. 1999, ApJ 515, 337
\bibitem[1994]{CR}Chiang J., Romani R.W. 1994, ApJ 436, 754
\bibitem[1965]{CL}Clark B.G. 1965, ApJ 142, 1398
\bibitem[1998]{CU}Cusumano G., Mineo T., Segreto A. et al. 1998, The Active
X-ray Sky (L. Scarsi et al. eds.), Nucl. phys. B (Proc. Suppl.) 69, 265
\bibitem[1996]{DH}Daugherty J.K., Harding A.K. 1996, ApJ 458, 278
\bibitem[1997]{EMM}Eastlund B.J., Miller B., Curtis Michel F. 1997, ApJ 483, 
     857
\bibitem[1995]{FR}Fierro J.M. 1995, Ph.D. Thesis, Stanford Univ.
\bibitem[1998]{FM}Fierro J.M., Michelson P.F., Nolan P.L. et al. 1998, 
ApJ 494, 734
\bibitem[1976]{FR}Fritz G., Meekins J.F., Chubb T.A. et al. 1976, ApJ 164, L55
\bibitem[1997]{FUK}Fukazawa Y., Ishida M., Ebisawa K. 1997, ASCA News 5, 1
\bibitem[1975]{GO}Gorenstein P. 1975, ApJ 198, 95
\bibitem[1998]{GU}Gull T.R., Lindler D.J., Crenshaw D.M. et al. 1998, 
ApJ 495, L51
\bibitem[1997]{MFM}Massaro E., Feroci M., Matt G. 1997, A\&AS 124, 123
\bibitem[1998]{MS}Massaro E., Feroci M., Costa E. et al. 1998 A\&A 338, 184
\bibitem[1997]{MIN}Mineo T., Cusumano G, Segreto A. et al. 1997, A\&A 327, L21
\bibitem[1997]{MT}Miyazaki J., Takahara F. 1997, MNRAS 290, 49
\bibitem[1995]{MU}Much R., Bennett K., Buccheri R. et al. 1995 A\&A 299, 435
\bibitem[1993]{PE}Percival J.W., Biggs J.D., Dolan J.F. et al. 1993, 
ApJ 407, 276
\bibitem[1997]{PAH}Pravdo S.H., Angelini L., Harding A.K. (PAH) 1997, ApJ 491, 808
\bibitem[1981]{PRS}Pravdo S.H., Serlemitsos.P.J. 1981, ApJ 246, 484
\bibitem[1995]{PS}Predehl P., Schmitt J.H.M.M. 1995 A\&A 293, 889
\bibitem[1985]{RL}Rieke G.H., Lebofsky M.J. 1985, ApJ 288, 618
\bibitem[1995]{RY}Romani R.W., Yadigaroglu I.-A. 1995, ApJ 438, 314
\bibitem[1988]{SM}Smith F.G., Jones D.H.P., Disck J.S.B. at al. 1988, 
MNRAS 233, 305
\bibitem[1994]{SD}Sturner S.J., Dermer C.D. 1994, ApJ 420, L79
\bibitem[1997]{TH}Thompson D.J., Harding A.K., Hermsen W. et al. 1997,
Proc. 4th Compton Symp. (C.D. Dermer, M.S. Strickman, J.D. Kurfess
eds), AIP Conf. Proc. 410, 39
\bibitem[1995]{UL}Ulmer M.P., Matz S.M., Grabelsky D.A. et al. 1995, ApJ 
     448, 356
\bibitem[1998]{WRH}Wang F.Y.-H., Ruderman M., Halpern J.P. et al. 1998 
     ApJ 498, 373 
\bibitem[1990]{ZB}Zombeck M 1990, Handbook of space astronomy and 
     astrophysics, Cambridge Univ. Press
\bibitem[1997]{ZR}Zhu T., Ruderman M. 1997, ApJ 478, 701
\end{thebibliography}
\end{document}